\documentclass{iopart}
\usepackage{iopams}

\def\xt{x(\tau)}
\def\xtp{x(\tau')}
\def\w{\omega}
\def\w0{\omega_{0}}
\def\wn{\omega_{n}}
\def\wk{\omega_{k}}
\def\bk{{\bf k}}
\def\bx{{\bf x}}
\def\br{{\mathbf r}}
\def\ak{a_{\bk\j}}
\def\akd{a_{\bk\j}^{\dagger}}

\def\ad{\langle a |}

\def\j{\mathit j}
\def\E{\mathbf{E}}
\def\m{\mathit m}

\usepackage{bm}
\usepackage{graphicx}

\usepackage{hyperref}
\hypersetup{colorlinks,
linkcolor=blue,
filecolor=green,
urlcolor=blue,
citecolor=blue}

\begin{document}
\title{Effects of a uniform acceleration on atom-field interactions}
\author{Jamir Marino$^{1,2}$, Antonio Noto$^{3,4}$, Roberto Passante$^3$, Lucia Rizzuto$^3$ and Salvatore Spagnolo$^3$}
\address{$^1$ SISSA International School for Advanced Studies, Via Bonomea 265, 34136 Trieste, Italy}
\address{$^2$ INFN, Istituto Nazionale di Fisica Nucleare, Sezione di Trieste, Italy}
\address{$^3$ Dipartimento di Fisica e Chimica, Universit\`{a} degli Studi di Palermo and CNISM, Via Archirafi 36, I-90123 Palermo, Italy}
\address{$^4$ Universit\'e Montpellier  2, Laboratoire Charles Coulomb UMR 5221- F-34095, Montpellier, France}
\ead{roberto.passante@unipa.it}
\begin{abstract}
We review some quantum electrodynamical effects related to the uniform acceleration of atoms in vacuum. After discussing the energy level shifts of a uniformly accelerated atom in vacuum, we investigate the  atom-wall Casimir-Polder force for accelerated atoms, and the van der Waals/Casimir-Polder interaction between two accelerated atoms.  The possibility of  detecting the Unruh effect through these phenomena is also discussed in detail.
\end{abstract}
\pacs{12.20.Ds, 42.50.Ct, 03.70.+k, 42.50.Lc}
\section{Introduction}
One of the most interesting phenomena in quantum field theory is the so-called Unruh effect \cite{Unruh76,CHM08}, stating that a uniformly accelerated detector in the Minkowski quantum vacuum experiences a thermal bath at a temperature proportional to its  acceleration. In qualitative terms, this phenomenon originates from  time-dependent  Doppler shifts of the field detected by the accelerated detector \cite{Milonni04}. Unfortunately, this effect is very tiny, and there is not yet an experimental evidence of it. In fact, an acceleration of the order of $10^{22}$ cm/s$^2$ would be necessary to obtain Unruh radiation corresponding to the temperature of 1 K \cite{CHM08}.
The question of the perception of the quantum vacuum in accelerated frames remains a widely debated problem.
In this paper we discuss and review some effects related to a uniform acceleration of atoms in the vacuum space, in the framework of quantum electrodynamics.
After considering the radiative level shifts of a uniformly accelerated atom in vacuum, we focus on the atom-wall Casimir-Polder interaction for an accelerated atom, as well as on the van der Waals/Casimir-Polder interaction between two accelerating atoms.
We are interested to investigate if  {\em  thermal} effects due to the acceleration, may produce observable effects in such physical systems. This is indeed expected because the Lamb-shift  and the Casimir-Polder interactions are directly related to vacuum field fluctuations \cite{Welton,CP48,PT93}.
This suggests the possibility of detecting the Unruh effect through a measurement of the Casimir-Polder interactions for atoms accelerating in the vacuum space.

The paper is organized as follows.
In Section 2, we review the Lamb-shift of a uniformly accelerated hydrogen atom in vacuum, in terms of vacuum fluctuations and radiation reaction field. Section 3 is devoted to the Casimir-Polder force between an accelerated atom and a conducting wall, both in the case of the scalar and the electromagnetic field. Finally, in Section 4 we investigate the atom-atom van der Waals/Casimir-Polder force between two accelerating atoms.

\section{Energy level shifts for an accelerated hydrogen atom}
\label{sec 2}

Let us  consider a hydrogen atom moving with uniform acceleration and interacting with the quantum electromagnetic field in the vacuum state. The Hamiltonian describing the atom-field interaction in the instantaneous inertial frame of the atom,  in the multipolar coupling scheme is \cite{Passante98,YZ06}

\begin{equation}
H(\tau)=H_{\rm A}(\tau)+H_{\rm F}(\tau)+H_{\rm I}(\tau) \label{eq:2.1} \, ,
\end{equation}
with
\begin{eqnarray}
&\ & H_{\rm A}(\tau)=\hbar \sum_n\wn\sigma_{nn}(\tau),\label{eq:2.2}\\
&\ & H_{\rm F}(\tau)=\hbar \sum_{\bk\j} \wk \akd\ak\frac{\rmd t}{\rmd\tau},
\label{eq:2.3}\\
&\ &H_{\rm I}(\tau)=-e\sum_{mn}\mathbf{r}_{mn}\cdot\E(\xt)\sigma_{mn}(\tau) \, ,
\label{eq: 2.4}
\end{eqnarray}
where $\tau$ is the proper time and $\sigma_{\ell m}=|\ell\rangle\langle m|$,  $|n\rangle$ being a complete set of atomic states with energy
$\omega_n$. ${\bf \mu}=e{\bf r}$ is the atomic electric dipole moment. Also, $x=(t,\bx)$ is the space-time coordinate of the atom and $\bk\j$ the wave vector ($j=1,2$ is the polarization index).
We are interested in investigating the energy level shifts of the uniformly accelerated atom. Exploiting the general procedure in \cite{DDC82,AM95}, we consider the contributions of vacuum fluctuations and radiation reaction field (indicated respectively with the subscripts \textit{vf} and \textit{rr}) to the atomic level shifts of the accelerated atom.
These quantities, at second order in $e$, are \cite{AM95}
\begin{eqnarray}
(\delta E_a)_{\rm vf}=- \frac{\rmi e^2}{\hbar}
\int_{\tau_0}^{\tau}\rmd\tau'C^{\rm F}_{\ell\m}(\xt,\xtp){(\chi^{\rm A}_{\ell\m})}_a(\tau,\tau'),
\label{eq:2.5}\\
(\delta E_a)_{\rm rr}=-\frac{\rmi e^2}{\hbar}
\int_{\tau_0}^{\tau}\rmd\tau'\chi^{\rm F}_{\ell\m}(\xt,\xtp){(C^{\rm A}_{\ell\m})}_a(\tau,\tau'),
\label{eq:2.6}
\end{eqnarray}
where $C_{\ell m}^{\rm F (A)}$ and $\chi_{\ell m}^{\rm F(A)}$ are the symmetric correlation function and the linear susceptibility of the field (atom), respectively.
Using the appropriate statistical functions of atom and field \cite{Takagi}, after some algebra,  it is obtained  \cite{Passante98}
\begin{eqnarray}
{(\delta E_a)}_{\rm vf}&=\frac{e^2}{3\pi c^3}\sum_b \left| \ad \br (0)|b \rangle \right|^2 \int_0^{\infty}\rmd\omega\,\omega^3 \left(1+\frac{a^2}{c^2\omega^2} \right)\nonumber \\
&\times \coth\left(\frac{\pi c\,\omega}{a}\right) P\left(\frac{1}{\omega+\omega_{ab}}-\frac{1}{\omega-\omega_{ab}}\right)\label{eq:2.7},\\
{(\delta E_a)}_{\rm rr}&= - \frac{e^2}{3\pi c^3}\sum_b \left| \ad \br (0)|b \rangle \right|^2 \int_0^{\infty}\rmd\omega\,\omega^3\nonumber \\
&\times P\left(\frac{1}{\omega+\omega_{ab}}+\frac{1}{\omega-\omega_{ab}}\right) \, ,
\label{eq:2.8}
\end{eqnarray}
where the index $a$ and the relative ket indicate a generic atomic state, $\omega_{ab}=\omega_a-\omega_b$, $a$ is the acceleration, and the limit $\tau-\tau_0$ to infinity has been taken.

We first note from (\ref{eq:2.8}) that the
radiation reaction contribution to the energy level shift does not depend on the atomic acceleration; it is identical to that obtained in the inertial case. This result is expected: in fact the field emitted by the atom propagates with the velocity of light, and the only moment it can act back on the atom is precisely the same time it is emitted. Thus, radiation reaction contribution is not influenced by the atomic motion. As we shall discuss in the next Section, the situation radically changes in the presence of boundaries, such as a reflecting mirror.
On the other hand, the contribution of vacuum fluctuations depends explicitly on the acceleration, through the presence of the {\em thermal} term $\coth(\pi c\,\omega/a)$, linked to the Unruh temperature $T=\hbar a/2\pi c k_{\rm B}$, and of an extra term proportional to $a^2$. This result indicates that the atomic acceleration induces observable effects in the energy shifts and in the observable Lamb shift.

The appearance of a nonthermal term proportional to $a^2$ is related to a similar term appearing in the correlation function of the electric field in the accelerated frame. It is possible to show that, for a ground-state hydrogen atom, thermal and non-thermal terms are comparable for $a\sim 10^{25}$ $\rm{cm/s^2}$  \cite{Passante98}. This is the also the typical acceleration required to detect the Unruh effect for such a system by measuring atomic level shifts.

\section{The atom-wall Casimir-Polder interaction for accelerated atoms}
\label{sec 3}

The same physical arguments given in the previous Section indicate that also the Casimir-Polder interaction between a uniformly accelerated atom and a perfectly reflecting plate could make manifest the Unruh effect.
Corrections to the atom-wall Casimir-Polder force due to the acceleration of the atom, have been calculated in the scalar field case \cite{Rizzuto07}.  It has been shown that such corrections are relevant only for accelerations of the order of $10^{24}$ cm/s$^2$, confirming the necessity of extremely high accelerations for a detection of the Unruh effect.
This calculation can be extended to the more realistic case of a uniformly accelerated hydrogen atom, interacting with the quantum electromagnetic field in the presence of a perfectly reflecting mirror \cite{RS09,ZY10}. Let us consider an atom moving with uniform acceleration in a direction parallel to the mirror. In analogy with the case of an atom at rest near a plate, the atom-wall Casimir-Polder interaction can be obtained considering only the $z$-dependent terms in the expression of the energy level shift. As before, we evaluate the contribution of vacuum fluctuations and of the radiation reaction field to the energy shift of the atomic level, in the presence of a conducting plate. After some lengthy algebra, it is found \cite{RS09}
\begin{eqnarray}
{(\delta E_a)}_{\rm{vf}}^{(\mathit b.c.)}&=-\frac{1}{8\pi^2 c^3}\sum_b\left(\mu_{\ell}^{ab}\mu_m^{ba}\right)\frac{1}{(2z_0)^3}P\int_0^{\infty}\rmd\omega
K_{\ell m}(\omega;z_0,a) \nonumber \\
&\times \coth\left(\frac{\pi c\,\omega}{a}\right)\left(\frac{1}{\omega+\omega_{ab}}-\frac{1}{\omega-\omega_{ab}}\right)
\label{eq:3.1}
\end{eqnarray}
and
\begin{eqnarray}
{(\delta E_a)}_{\rm{rr}}^{(b.c.)}&=
\frac{1}{8\pi^2 c^3}\sum_b\left(\mu_{\ell}^{ab}\mu_m^{ba}\right)\frac{1}{(2z_0)^3}P\int_0^{\infty}\rmd\omega \nonumber \\
&\times K_{\ell m}(\omega;z_0,a)
\left(\frac{1}{\omega+\omega_{ab}}+\frac{1}{\omega-\omega_{ab}}\right) \, ,
\label{eq:3.2}
\end{eqnarray}
where the superscript (b.c.) stands for \textit{boundary conditions}, $z_0$ is the atom-wall distance and $K_{\ell m}(\omega;z_0,a)$ is a function containing a combination of sinusoidal functions, which takes into account the presence of the conducting plate \cite{RS09}.

We now briefly comment the results obtained. Equation (\ref{eq:3.1})  clearly shows that the contribution of vacuum fluctuations contains not only a thermal correction due to the factor $\coth(\pi c\,\omega/a)$,  but also an extra term proportional to the function  $K_{\ell m}(\omega;z_0,a)$.  This function modulates the Casimir-Polder interaction as a function of the atom-plate distance $z_0$ and of the atomic acceleration $a$.
On the other hand, Equation (\ref{eq:3.2}) shows that the radiation reaction term is sensitive to the atomic acceleration. This behavior is not surprising. When a boundary is present, the field emitted by the atom can act back on the atom after a reflection on the conducting plate. Since the atom accelerates, in the time-interval between the emission and the subsequent absorption of the reflected field, the atom has moved from its position of a distance depending on its acceleration. This gives rise to a dependence of the radiation reaction contribution on the atomic acceleration.
The expression for the total atom-wall Casimir-Polder interaction for the accelerated atom, is obtained by summing (\ref{eq:3.1}) and  (\ref{eq:3.2}).
It is easy to show that in order to reveal effects of acceleration on the atom-wall Casimir-Polder interaction, accelerations of the order of $10^{24}$cm/s$^2$ are necessary, as in the case of the energy shift of an atom in the unbounded space. This makes very difficult to observe the effects of the acceleration through the Lamb shift or the atom-wall Casimir-Polder interactions. As we shall show in the next Section, the situation seems different when we consider the van der Waals/Casimir-Polder interaction between two accelerated atoms.

\section {van der Waals/Casimir-Polder interaction energy between two accelerated atoms}

We now discuss the effect of the uniform acceleration on the van der Waals interaction  (both in the non-retarded and in the Casimir-Polder regime)  between two atoms in their ground state moving parallel to each other, with the same uniform acceleration perpendicular to their distance.  A first approach to this problem can be done by extending a procedure developed in  \cite{GW} for the Casimir-Polder interaction at finite temperature, to the case of accelerated atoms, on the basis of the equivalence between temperature and acceleration expressed by the Unruh effect. Following this method one finds, in the low acceleration limit ($a\ll\omega_0 c$), a qualitative change of the distance dependence of the interaction energy due to the acceleration in the far zone, and a correction in the near zone \cite{MP10}. This result suggests that investigating the behavior of the van der Waals force could give evidence of the Unruh effect.

An alternative and more fundamental approach to this problem is obtained by generalizing a heuristic model for the van der Waals interactions, already used for atoms at rest \cite{PT93,PPR03,CP97,Salam09} and in the presence of boundary conditions \cite{SPR06}, to the case of accelerating atoms \cite{NP13}.  In such model the interaction energy arises from the dipolar interaction between the (instantaneous) oscillating dipole moments of the atoms, induced and spatially correlated by the zero-point electromagnetic field fluctuations. The dipole fields are treated classically while the quantum nature of the radiation is included in the spatial correlations of the electric field related to vacuum fluctuations.

We start calculating the electric and the magnetic field of an accelerating dipole (atom $A$) in the position of the other dipole (atom $B$) at the retarded time \cite{PT01}. These fields are then Lorentz-transformed to the co-moving frame of the atoms, that is a locally inertial frame.
The interaction of the field emitted by atom A with atom B, related to the dipole moments correlation and thus to the spatial correlations of vacuum field fluctuations, is then calculated in this frame and expressed in terms of physical quantities in the laboratory frame.
We finally get the following expression of the interaction energy, at time $t$ (the atoms are assumed at rest at $t=0$) \cite{NP13}

\begin{eqnarray}
\langle \Delta \tilde{E}\rangle  &= \Delta E^{\rm r}+\frac{a^2 t}{2 c^3}\frac{\hbar c}{\pi R^3} \int _0^\infty \alpha (A; \rmi u)\alpha(B; \rmi u) \nonumber \\
&\times \left( 3 + \frac{4}{uR} +\frac{2}{u^2R^2}\right)u^2\,\exp[-2uR]\, \rmd u\, \nonumber \\
&+\frac{a^2t^2}{6c^2}\frac{\hbar c}{\pi R^2}\,\int _0^\infty \alpha (A; \rmi u)\alpha(B; \rmi u) \, \left( -1+\frac{4}{uR}\right.\nonumber \\
&+\frac{8}{u^2R^2}\left.+\frac{8}{u^3R^3}+\frac{4}{u^4R^4}\right) u^4\,\exp[-2uR]\,\rmd u \, ,
\label{eq:32}
\end{eqnarray}
where
\begin{eqnarray}
\Delta E^{\rm r} &= - \frac{\hbar c}{\pi R^2} \int _0^\infty \alpha (A; \rmi u)\alpha(B; \rmi u) \left[ 1 + \frac{2}{uR} + \frac{5}{u^2R^2}\right.\nonumber \\
&+\left.\frac{6}{u^3R^3}+\frac{3}{u^4R^4}\right] u^4 \exp[-2uR]\, \rmd u
\label{eq:33}
\end{eqnarray}
is the well-known van der Waals potential energy for two atoms at rest \cite{CP48,PT93}. $\alpha$ is the atomic dynamical polarizability and $R$ the interatomic distance.

From (\ref{eq:32}) it is possible to show that in the near zone, an acceleration-dependent term proportional to $R^{-5}$ adds to the usual $R^{-6}$ term,  while in the far zone a term proportional to $R^{-6}$ adds to the usual $R^{-7}$ behavior.
There are also acceleration-dependent corrections to the usual $R^{-6}$ (near zone) and $R^{-7}$ (far zone) terms. These corrections depend on the product $at$; thus, by taking sufficiently long times, significant changes to the interaction energy could be observed even for reasonable values of the acceleration.
For example, if we consider $a^2t^2/c^2\simeq 0.2$,  the correction term is about 10\% in near zone and 1\% in far zone; taking $t\simeq 10^{-3}$ s, we obtain $a\sim 10^{13}$ cm/s$ ^2$, an acceleration several orders of magnitude smaller than that necessary to observe the Unruh effect in the cases discussed in the previous Sections.

A different formal approach to the problem can be based on an extension of the method used in Sections \ref{sec 2} and \ref{sec 3} to the fourth-order in the coupling, allowing separation of vacuum fluctuations and radiation reaction contributions to the Casimir-Polder interaction between two accelerated atoms. For atoms interacting with a scalar field, we obtain, after lengthy algebra, the following expression for the contribution of vacuum fluctuations to distance-dependent energy level shifts \cite{MNP13}
\begin{eqnarray}
{(\Delta E)}_{\rm vf} &=4\rmi\mu^4\int_{\tau_0}^\tau \rmd\tau'\int_{\tau_0}^{\tau'} \rmd\tau''\int_{\tau_0}^{\tau''} \rmd\tau'''
C^{\rm F}\left(\phi^{\rm f}(x_{\rm A}(\tau)),\phi^{\rm f}(x_{\rm B}(\tau'''))\right)
\nonumber \\
&\times  \chi^{\rm F}\left(\phi^{\rm f}(x_{\rm A}(\tau')),\phi^{\rm f}(x_{\rm B}(\tau''))\right)
\chi_b^{\rm B}(\tau'',\tau''')\chi_a^{\rm A}(\tau,\tau') \, ,
\end{eqnarray}
where $a$ and $b$ are respectively the states of atom $\rm A$ and atom B, $\mu$ is the coupling constant, $\tau$ the proper time and $\phi^{\rm f}$ the free part of the scalar field. $C^{\rm F}$ and $\chi^{\rm F}$ indicate the symmetrical correlation function and susceptivity of the field, respectively, while $\chi^{\rm A(B)}$ is the susceptivity of atom A (B).
An analogous expression can be obtained for the radiation reaction contribution. From the distance-dependent part of the energy shift $\delta E_a^{\rm A}$ obtained with this method,  using the appropriate statistical functions of field and atoms, we can obtain the van der Waals/Casimir-Polder interaction energy. This method has also the advantage that it can be used in several different cases (atoms at rest, atoms in a thermal field or moving atoms, for example), by
substituting the appropriate statistical functions.

\section {Conclusions} We have reviewed several different physical effects related to the uniform acceleration of atoms in the quantum vacuum, in the framework of quantum electrodynamics. Specifically, after discussing the energy level shifts of a uniformly accelerated atom in vacuum, we have investigated the  atom-wall Casimir-Polder force for an accelerated atom,  as well as the van der Waals/Casimir-Polder interaction between two accelerated atoms, using different approaches. We have shown that the atomic acceleration modifies both the Lamb shift and the Casimir-Polder or van der Waals interactions. In particular, our results suggest that significant changes to the interaction energy between the two accelerated atoms could be observed even for reasonable values of the acceleration, contrary to other known manifestations of the Unruh effect.
\section{Acknoledgments}
The authors gratefully acknowledge financial support by the Julian Schwinger Foundation, by Ministero dell'Istruzione, dell'Universit\`{a} e della Ricerca and by Comitato Regionale di Ricerche Nucleari e di Struttura della Materia.


\section*{References}
\begin{thebibliography}{50}
\bibitem{Unruh76} Unruh W G 1976 \PR D {\bf 14} 870
\bibitem{CHM08} Crispino L C B, Higuchi A and Matsas G E A 2008 \RMP {\bf 80} 787
\bibitem{Milonni04} Alsing P M and Milonni P W 2004 {\it Am J. Phys.} {\bf 72} 1524
\bibitem{Welton} Welton T A 1948 \PR {\bf 74} 1157
\bibitem{CP48} Casimir H B G and Polder D 1948 \PR {\bf 73}, 360
\bibitem{PT93} Power E A and Thirunamachandran T 1993 \PR A {\bf 48} 4761
\bibitem{Passante98} Passante R 1998 \PR A \textbf{57} 1590
\bibitem{YZ06} Yu H and Zhu Z 2006 \PR D \textbf {74} 044032
\bibitem{DDC82} Dalibard J, Dupont-Roc J and Cohen-Tannoudji C 1982 \JP {\bf 43} 1617; Dalibard J, Dupont-Roc J and Cohen-Tannoudji C 1984 \JP {\bf 45} 637
\bibitem{AM95} Audretsch J and Muller R 1995 \PR A \textbf{52} 629
\bibitem{Takagi} Takagi S 1988 {\it Prog. Theor. Phys. Suppl.} \textbf {88} 1
\bibitem{Rizzuto07} Rizzuto L 2007 \PR A {\bf 76} 062114
\bibitem{RS09} Rizzuto L and Spagnolo S 2009 \PR A {\bf 79} 062110
\bibitem{ZY10}  Zhu Z and Yu H 2010 \PR A {\bf 82} 042108
\bibitem{GW} Goedecke G H and Wood R 1999 \PR A {\bf 60} 2577
\bibitem{MP10} Marino J and Passante R 2010 {\it Quantum Field Theory under the Influence of External Conditions (QFEXT09)} edited by Milton K A and Bordag M (Singapore: World Scientific) p~328
\bibitem{PPR03} Passante R, Persico F and Rizzuto L 2003 \PL A {\bf 316} 29
\bibitem{CP97} Cirone M and Passante R 1997 {\it J. Phys.} B {\bf 30} 5579
\bibitem{Salam09} Salam A 2009 {\it J.Phys.: Conf. Ser.} {\bf 161} 012040
\bibitem{SPR06} Spagnolo S, Passante R and Rizzuto L 2006 \PR A {\bf 73} 062117
\bibitem{NP13} Noto A and Passante R 2013 \PR D {\bf 88} 025041
\bibitem{PT01} Power E A and Thirunamachandran T 2001 {\it Proc. R. Soc. Lond. A 457} 2757
\bibitem{MNP13} Marino J, Noto A and Passante R, 2013 {\it in preparation}
\end{thebibliography}
\end{document}